\documentclass[aps,prstab,preprint,showkeys,groupedaddress]{revtex4-1}
\usepackage{graphicx}
\usepackage{subcaption}
\usepackage{stmaryrd}
\usepackage{relsize}
\usepackage{amsmath}
\usepackage{color}
\usepackage[english]{babel}
\usepackage[nottoc]{tocbibind}
\usepackage{epstopdf}

\begin{document}
\title{Closed orbit correction at synchrotrons for symmetric and near-symmetric lattices}
\author{S. H. Mirza}
\email[Corresponding author: ]{s.h.mirza@gsi.de}
\affiliation{GSI Helmholtzzentrum f\"ur Schwerionenforschung GmbH, Darmstadt, Germany}
\affiliation{Technische Universit\"at Darmstadt, Darmstadt, Germany}
\author{R. Singh}
\affiliation{GSI Helmholtzzentrum f\"ur Schwerionenforschung GmbH, Darmstadt, Germany}
\author{P. Forck}
\affiliation{GSI Helmholtzzentrum f\"ur Schwerionenforschung GmbH, Darmstadt, Germany}
\author{H. Klingbeil}
\affiliation{GSI Helmholtzzentrum f\"ur Schwerionenforschung GmbH, Darmstadt, Germany}
\affiliation{Technische Universit\"at Darmstadt, Darmstadt, Germany}
\begin{abstract}
This contribution compiles the benefits of lattice symmetry in the context of closed orbit correction. A symmetric arrangement of BPMs and correctors results in structured orbit response matrices of Circulant or block Circulant type. These forms of matrices provide favorable properties in terms of computational complexity, information compression and interpretation of mathematical vector spaces of BPMs and correctors. For broken symmetries, a nearest-Circulant approximation is introduced and the practical advantages of symmetry exploitation are demonstrated with the help of simulations and experiments in the context of FAIR synchrotrons. 
\end{abstract}


\pacs{}

\maketitle
\section{Introduction}
The closed orbit correction has been an integral part of synchrotron and storage ring in light sources as well as in hadron machines for stable beam operations~\cite{gen,spear,LHC}. Closed orbit correction methods are typically classified as ``local'' or ``global'' given the spatial extent of their correction. Local bumps generated by three to four correctors are utilized for orbit correction in a localized region of a synchrotron while global correction methods rely on the effect of each corrector throughout the synchrotron. The global effect of a single dipole kick $\theta_{c}$ located at a longitudinal location $s_{1}$ is described by the solution of the Hill's equation~\cite{Sands,svd1} 
\begin{equation}\label{eq1}
    z(s-s_{1})=\theta_{c} \frac{\sqrt{\beta(s_{1})\beta(s)}}{2\sin(\pi Q_{z})}{\cos \left(Q_{z} \pi-|\mu(s_{1})-\mu(s)|\right)}
\end{equation}
where $s-s_{1}$ is the longitudinal separation from the source of dipolar kick and $z$ is the transverse orbit position in either plane (i.e. $z$ is either $x$ or $y$). $\beta$ and $\mu$ denote the lattice beta function and phase advance, respectively and $Q_{z}$ is the coherent betatron tune in either plane. For a finite number of beam position monitors (BPMs) and correctors, Eq.~\ref{eq1} takes the shape of a matrix referred to as the orbit response matrix (ORM) such that:
\begin{equation}\label{eq1s}
     z=\textbf{R}\Theta.
\end{equation}
 where $\Theta$ is the corrector settings vector and $z$ is the beam positions vector at the BPM locations. In nutshell, the main concept of global correction is to calculate the corrector strengths that can counteract the existing dipolar field errors such that the orbit distortion measured with the BPMs is minimized. \\
Historically, four distinct methods have served the global orbit correction which include sliding bump method~\cite{lbmp}, MICADO~\cite{micado}, harmonic correction~\cite{rha} and singular value decomposition (SVD)~\cite{svd1}. A variant of the SVD type correction referred to as eigenvalue decomposition has also been reported~\cite{eigval,Fried}. Sliding bump method involved forming independent local bumps in order to achieve the required positions at the BPM locations and has been phased out in usage. MICADO also referred to as orthogonal matching pursuit (OMP) in the signal processing literature~\cite{omp1,omp2} was devised to find out some most effective correctors for minimizing the orbit distortion and has robustness and computational issues. Harmonic correction was the first method to discuss the notion of mode-based correction by means of a decomposition of the perturbed orbit into Fourier harmonics which can be corrected individually. However, the validity and efficacy of this method for non-periodic lattices was not explored.
In literature, it seems to have been used only when the correction is intended for few specific spatial modes of the perturbed orbit, e.g. modes around the coherent tune frequency. Singular value decomposition (SVD) is a generalized technique based upon diagonalization and inversion of the matrices and superseding all the above-mentioned methods, has become the de-facto algorithm for orbit correction. The SVD of a real-valued matrix \textbf{R} is given as~\cite{svd} 
\begin{equation}\label{eq1_1}
\begin{split}
    \textbf{R}= \textbf{U}\textbf{S}\textbf{V}^{\text{T}},
    \end{split}
\end{equation}
where \textbf{U} and \textbf{V} are the left and right orthogonal matrices and \textbf{S} is the diagonal matrix whose diagonal entries are called singular values. Like harmonic analysis, SVD also provides the liberty of mode-by-mode orbit correction on top of mode-truncation and matrix inversion using a transformation of perturbed orbit vector $z$ and corrector settings vector $\Theta$ into the mode-space as~\cite{svd1} 
\begin{equation}\label{eq2}
\begin{split}
    \Bar{z}= \textbf{S}\Bar{\Theta},
    \end{split}
\end{equation}
where $\Bar{z} = \textbf{U}^{\text{T}}z$ and $\Bar{\Theta} = \textbf{V}^{\text{T}}\Theta$ are the BPM and corrector vectors in the transformed mode-space, respectively. The solution of Eq.~\ref{eq2} gives the required corrector settings for a given perturbed orbit.
SVD also has some limitations particularly when dealing with uncertainty in the process model as it is a numerical technique and there is no apparent analytic way of associating uncertainties in the lattice parameters to the singular values~\cite{sandira2}. Moreover, a lack of physical interpretation of SVD modes, their mutual phase relationship and dependence on singular values makes the uncertainty modelling complicated~\cite{Mirza,sandira1}. The inter-dependence between \textbf{U}, \textbf{S} and \textbf{V} matrices also poses a special challenge for systems where matrices need to be updated during orbit correction i.e. on the acceleration ramp in synchrotrons~\cite{ipac18}.\\
In this paper, we present a one dimensional Discrete Fourier Transform (DFT)-based diagonalization and inversion of the ORM for symmetric lattices. The technique is based upon the exploitation of Circulant symmetry in the lattice and provides computational benefits, information compression into a diagonal matrix only and physical interpretation of the ORM mode-space. This method serves as the transition between previously discussed harmonic analysis and SVD with an exact equivalence for symmetric lattices. Further, a nearest-Circulant extension is discussed for broken symmetries making most of the ideas discussed for symmetric matrices applicable to those of non-symmetric lattices.\\
In the next section, the DFT-based decomposition and inversion of the ORMs for full symmetry, block symmetry and broken symmetry cases are shown. All these cases are exemplified by the FAIR synchrotrons. Following that, the practical benefits of the technique are demonstrated by simulations and measurements at the SIS18 synchrotron of GSI.

\section{Symmetry in response matrix}
 Equations~\ref{eq1} and~\ref{eq1s} represent the orbit response matrix and show that it contains one global parameter, i.e. the betatron tune, while two local parameters, the beta function and phase advance, depend solely on the BPM and corrector locations. 
 The global parameter determines the general criterion for the number of BPMs and correctors for an effective orbit correction. In order to ensure that modes up to twice the coherent betatron frequencies can be corrected, at least 4 BPMs and correctors are required per betatron oscillation. The placement of BPMs and correctors is chosen at locations of higher beta function values in order to enhance the sensitivity of the closed orbit correction.  However during the design, the symmetric arrangement of BPMs and correctors is usually neither given the importance it deserves nor exploited even if it exists. The only notable discussion of a symmetric layout for BPMs and correctors was found in~\cite{Fried} but was limited to finding eigenvalues for eigenvector-based orbit correction. Any discussions towards the matrix inversion, relations to SVD and application to broken symmetry were not made. These issues will be discussed in this report. In this section, we will discuss two kinds of ORM symmetries which exist in two different synchrotrons of the FAIR project, the SIS18 and SIS100. We use them as practical examples and to extend our findings to broken symmetries.
\subsection{Circulant symmetry of SIS18 vertical ORM}\label{seccrc}
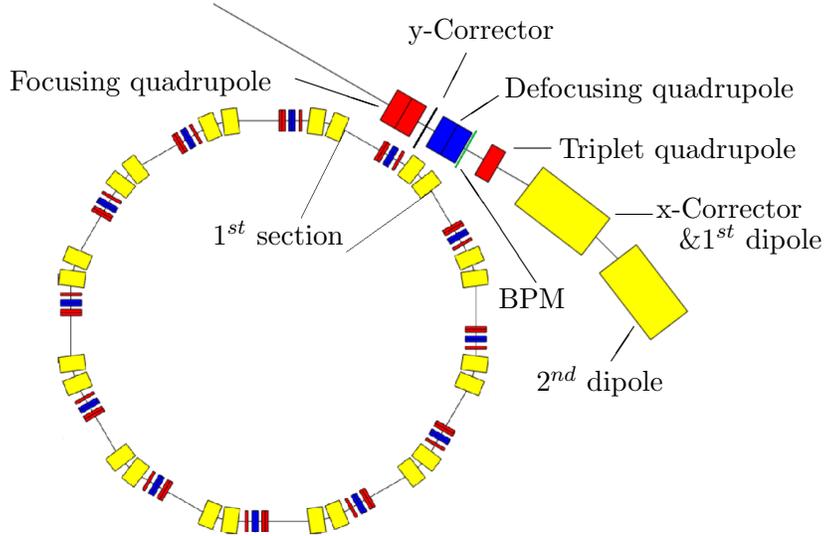
\begin{figure}
\centering
\begin{picture}(200,200)
\setlength{\unitlength}{1cm}
\put(0,0){\includegraphics[width=8.6 cm]{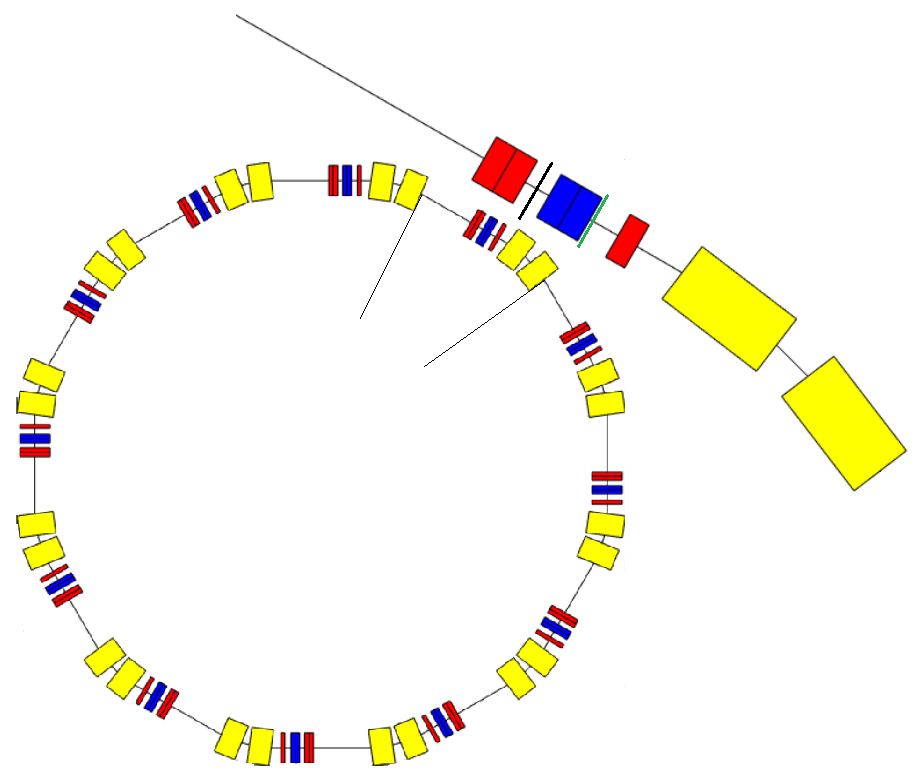}}
\put(4.8,6.68){y-Corrector}
\put(5.27,5.9){\line(1,2){0.3}}
\put(6.08,5.9){Defocusing quadrupole}
\put(5.5,5.6){\line(3,2){0.5}}
\put(6.8,5.1){Triplet quadrupole}
\put(6.2,5.23){\line(1,0){0.5}}
\put(8.1,4.3){x-Corrector}
\put(7.55,4.35){\line(1,0){0.5}}
\put(8.4,3.9){$\& 1^{st}$ dipole}
\put(-0.5,6){Focusing quadrupole}
\put(3.3,6.15){\line(3,-1){1.1}}
\put(6,3.1){BPM}
\put(6.5,3.45){\line(-2,3){1}}
\put(6.5,2.0){$2^{nd}$ dipole}
\put(7.5,2.4){\line(2,3){0.3}}
\put(2.2,3.95){$1^{st}$ section}
\end{picture}
    \caption{Schematic of the lattice of the SIS18 synchrotron. The first section has been magnified on top right and the components are labelled. The horizontal correctors are located at the $1^{st}$ dipole as extra windings in all sections except in the $4^{th}$ and $6^{th}$ sections where they reside on the second dipoles (not labelled in the figure). \label{fig1}}
\end{figure}

The SIS18 is a $216.72$ m long synchrotron and has a 12-fold symmetric lattice as shown in Fig.~\ref{fig1}. It comprises in the vertical plane of one BPM and one corrector each per section placed at the same location. In the horizontal plane, only two correctors violate this symmetry which will be discussed later in the context of broken symmetry.
The symmetric arrangement of BPMs and correctors results in an equal phase advance between adjacent BPMs 
with same beta function 
for all BPMs. The correctors also have a similar arrangement which gives a special shape to the ORM called a Circulant matrix in which each row and column is a cyclic permutation of the previous row and column, respectively, abbreviated as $circ$. The theory of Circulant matrices is well established in literature~\cite{circulant} and one out of many useful properties of Circulant matrices, in the context of ORMs, is their diagonalization and decomposition by the DFT of only one row or column. If \textbf{R}$_C$ is a square Circulant matrix of dimension $n$, it can be written as a sum of fundamental permutation matrices $\pi_{n}^i$ (see appendix A), each of which is convoluted by one element $r_{i}$ of the first row or column of \textbf{R}$_C$ as 
\begin{equation}
    \label{eq4}
    \textbf{R}_C = r_{0}\pi_{n}^{0}+r_1\pi_{n}^{1}+ ....... r_{n-1} \pi_{n}^{n-1}=\sum_{i=0}^{n-1} (\pi_n^{i}r_{i})
\end{equation}
where $i$ is the order of the permutation matrix. As an example, Eq.~\ref{eq3} shows the SIS18 vertical ORM calculated by MAD-X~\cite{madx} for triplet optics in the units of $mm/mrad$:
\begin{equation}
    \setcounter{MaxMatrixCols}{20}
    \label{eq3}
    \textbf{R}_{y}=\setlength\arraycolsep{1.5pt} circ
\begin{bmatrix}
\hspace{1mm}8.10&\hspace{1mm}8.90&-8.20&-6.40&\hspace{1mm}10.2&\hspace{1mm}3.30&-11.2&\hspace{1mm}0.10&\hspace{1mm}11.2&-3.50&-10.1&\hspace{1mm}6.60
\end{bmatrix}
\end{equation}
A circulant matrix can be decomposed as 
\begin{equation}
    \label{eq6}
    \textbf{R}_C = \textbf{F}^{*}\mathbf{\Lambda}\textbf{F} 
\end{equation}
where \textbf{F} is a standard Fourier matrix  which is identical for all Circulants of same size with elements given as~\cite{DFTmatrix} 
\begin{equation}
    \label{eq7}
    \centering
    \begin{split}
    (\textbf{F})_{f,i}= \frac{1}{\sqrt{n}}e^{j2 \pi fi/n}=\frac{1}{\sqrt{n}}\left(\cos\left(\frac{2 \pi fi}{n}\right)+j\sin\left(\frac{2 \pi fi}{n}\right)\right)
    \end{split}
\end{equation}
for $i,f \in [0,..,n-1]$ where $i$ represents the sampling points, $f$ is the discrete frequency of each Fourier mode (column of \textbf{F}), $j$ is the imaginary unit and $n$ is the size of the square Circulant matrix. $\mathbf{\Lambda}$ is a diagonal matrix containing the discrete Fourier coefficients $\sigma_f$ of the first row or column of \textbf{R}$_C$ on its diagonal positions, which are given as  
\begin{equation}
    \label{eq9}
    \centering
    \sigma_f = \text{Re}\{\sigma_{f}\} +j \text{Im}\{\sigma_{f}\}= \sum_{i=0}^{n-1} r_i e^{-j2 \pi fi/n}
\end{equation}
In the case of an ORM, $n$ is the total number of BPMs or correctors while the columns of matrix \textbf{F} represent the mode space of BPM and correctors comprised of pure sine and cosine functions. In this way, a DFT-based decomposition gives a physical interpretation to the mode-space of an ORM and would be equivalent to the harmonic analysis. 
The inverse of \textbf{R}$_C$ can be written as 
\begin{equation}
    \label{eq8}
    \centering
    \textbf{R}_C^{-1}=\textbf{F}^{*}\mathbf{\Lambda}^{-1}\textbf{F}
\end{equation}
where $\mathbf{\Lambda}^{-1}$ is the diagonal matrix having inverses of Fourier coefficients at its diagonal positions. 

\subsection{Equivalence of SVD and DFT for Circulant symmetry}\label{eqv}
For the general matrices, there is no analytic information available to interpolate the SVD modes between discrete elements of \textbf{U} and \textbf{V} matrices, as SVD is a numerical technique and it is free to choose any mode structure in order to satisfy the orthogonality of the \textbf{U} and \textbf{V} matrices. 
In case of Circulant matrices, there exists an equivalence between SVD and DFT by introducing the discrete Hartley transform matrix $\textbf{H}$. Following equation~\ref{eq6} and using the theorem $4.1$ of~\cite{Karner} which states that the SVD of a Circulant matrix can be written as
\begin{equation}
    \label{eq41}
    \centering
    \textbf{R}_{C} = \left(\textbf{H(F)}\text{Re}\{\mathbf{\Sigma}\}-\textbf{H}(\Bar{\textbf{F}})\text{Im}\{\mathbf{\Sigma}\}\right)|\mathbf{\Lambda}|\textbf{H}(\textbf{F})
\end{equation}
where
\begin{equation}
    \label{eq42}
    \centering
    \begin{split}
    \textbf{H(F)}=\text{Re}\{\textbf{F}\}+\text{Im}\{\textbf{F}\}\\
    \textbf{H}(\Bar{\textbf{F}})=\text{Re}\{\textbf{F}\}-\text{Im}\{\textbf{F}\}
    \end{split}
\end{equation}
The diagonal matrix $\mathbf{\Lambda}$ of the DFT-based decomposition can be written as 
\begin{equation}
    \label{eq43}
    \centering
    \mathbf{\Lambda}=\mathbf{\Sigma}|\mathbf{\Lambda}|
\end{equation}
with $\mathbf{\Sigma}$ and $|\mathbf{\Lambda}|$ being the diagonal matrices containing only the phases $\phi_{di}$ and magnitudes of each Fourier coefficient, respectively. The matrices \textbf{U}, \textbf{S} and \textbf{V} are calculated below by solving the right hand side of Eq.~\ref{eq41} element-wise. The last term can be solved as 
\begin{eqnarray}
    \label{eq44}
    \centering
         (\textbf{H(F)})_{f,i}&=&\frac{1}{\sqrt{n}}\left(\cos\left(\frac{2 \pi  \nonumber fi}{n}\right)+\sin\left(\frac{2 \pi fi}{n}\right)\right)\\ \nonumber
         &=&\sqrt{\frac{2}{n}}\left(\cos\left(\frac{2 \pi fi}{n}+\phi\right)\right) \nonumber\\ &=&(\textbf{V}^{\text{T}})_{f,i}   
\end{eqnarray}
where $\phi= -\frac{\pi}{4}$. The singular values of the SVD matrix \textbf{S} are the moduli of the Fourier coefficients of DFT diagonal matrix $\mathbf{\Lambda}$:
\begin{equation}
    \label{eq45}
    \centering
    \textbf{S}=|\mathbf{\Lambda}|
\end{equation}
Similarly, the \textbf{U} matrix is equal to the following first part of the right hand side of Eq.~\ref{eq41} as 
\begin{eqnarray}
    \label{eq45a}
    \centering
         &&\left(\textbf{H(F)}\text{Re}\{\mathbf{\Sigma}\}-\textbf{H}(\Bar{\textbf{F}})  \nonumber \text{Im}\{\mathbf{\Sigma}\right)_{f,i}\\  \nonumber
         &=&\sqrt{\frac{1}{n}}\left(\cos\left(\frac{2 \pi fi}{n}\right)+\sin\left(\frac{2 \pi  \nonumber fi}{n}\right)\right)\cos\left(\phi_{di}\right)-\frac{1}{\sqrt{n}}\left(\cos\left(\frac{2 \pi \nonumber  fi}{n}\right)-\sin\left(\frac{2 \pi fi}{n}\right)\right)\sin\left(\phi_{di}\right)\\  \nonumber
         &=&\sqrt{\frac{1}{n}}\left(\cos\left(\frac{2 \pi fi}{n}-\phi_{di}\right)+\sin\left(\frac{2 \pi \nonumber fi}{n}-\phi_{di}\right)\right)\\ \nonumber
         &=&\sqrt{\frac{2}{n}}\left(\cos\left(\frac{2 \pi fi}{n}-\phi_{di}-\frac{\pi}{4}\right)\right)\\
         &=&(\textbf{U})_{f,i}   
\end{eqnarray}
Combining equations~\ref{eq44},~\ref{eq45} and~\ref{eq45a}, equation~\ref{eq41} can be written as
\begin{equation}\label{eq45b}
\begin{split}
    \textbf{R}_C= \textbf{U}\textbf{S}\textbf{V}^{\text{T}}
    \end{split}
\end{equation}
This conversion shows a significant difference between the two techniques in terms of information spread. SVD distributes the information in all three matrices while DFT-based decomposition compresses all the information into one diagonal matrix.
SVD-like mode truncation is also possible here by removing the Fourier coefficients of absolute values below a certain threshold along with corresponding columns of \textbf{F} thus preserving the main benefit of SVD while adding the benefits of harmonic analysis.
\subsection{Broken symmetry and nearest-Circulant approximation}\label{secnc}
\begin{figure}[h!]
\centering
\includegraphics[width=8.3 cm]{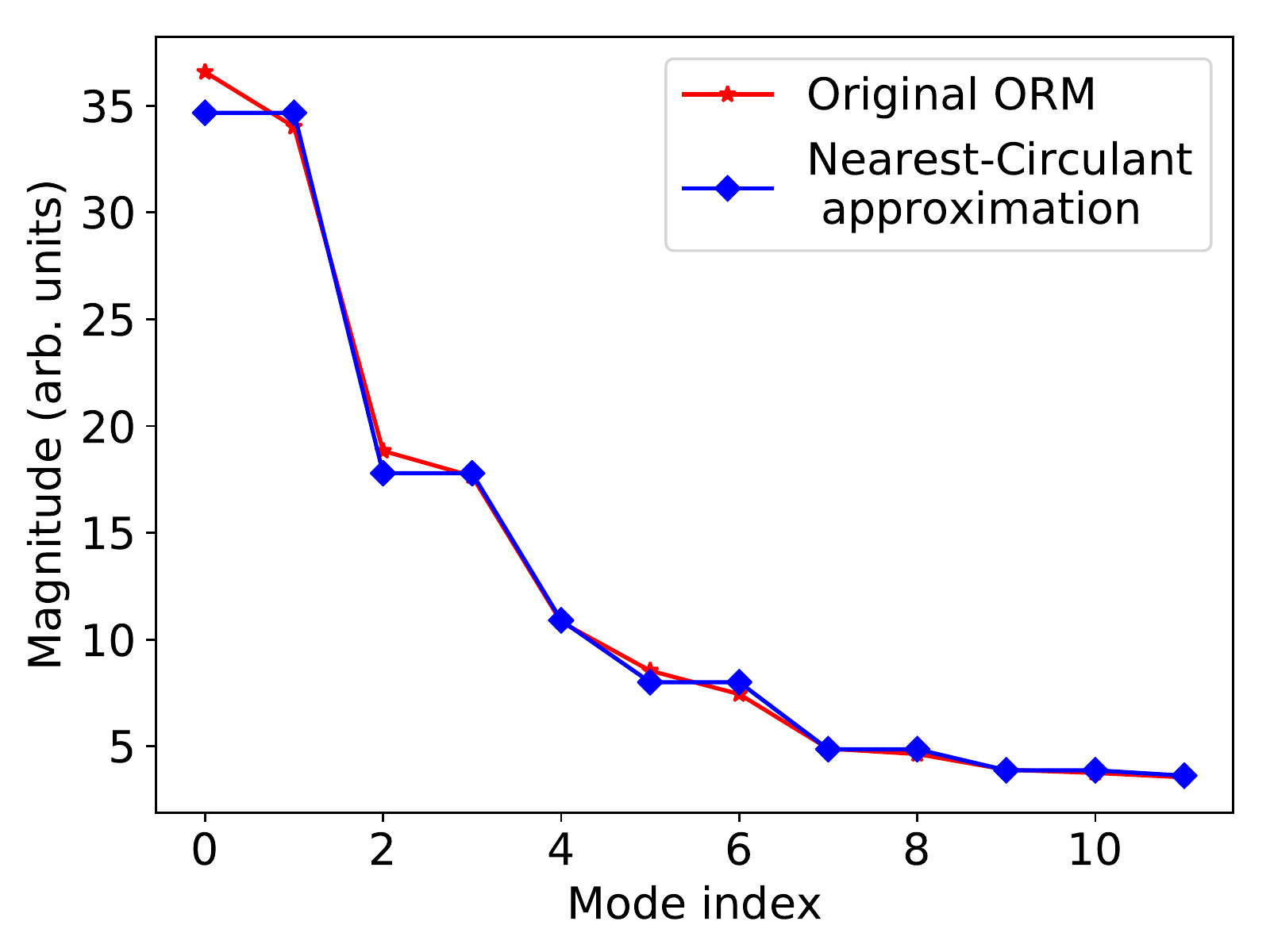}
\caption{Comparison of singular values of SIS18 ORM and its nearest-Circulant approximation in the $x-$plane. The abscissa is the index of singular values of the matrix $\textbf{S}$ while the ordinate is the magnitude of the singular values.\label{fig14}}
\end{figure}

\begin{figure}[h!]
\centering
\includegraphics[width=8.6 cm]{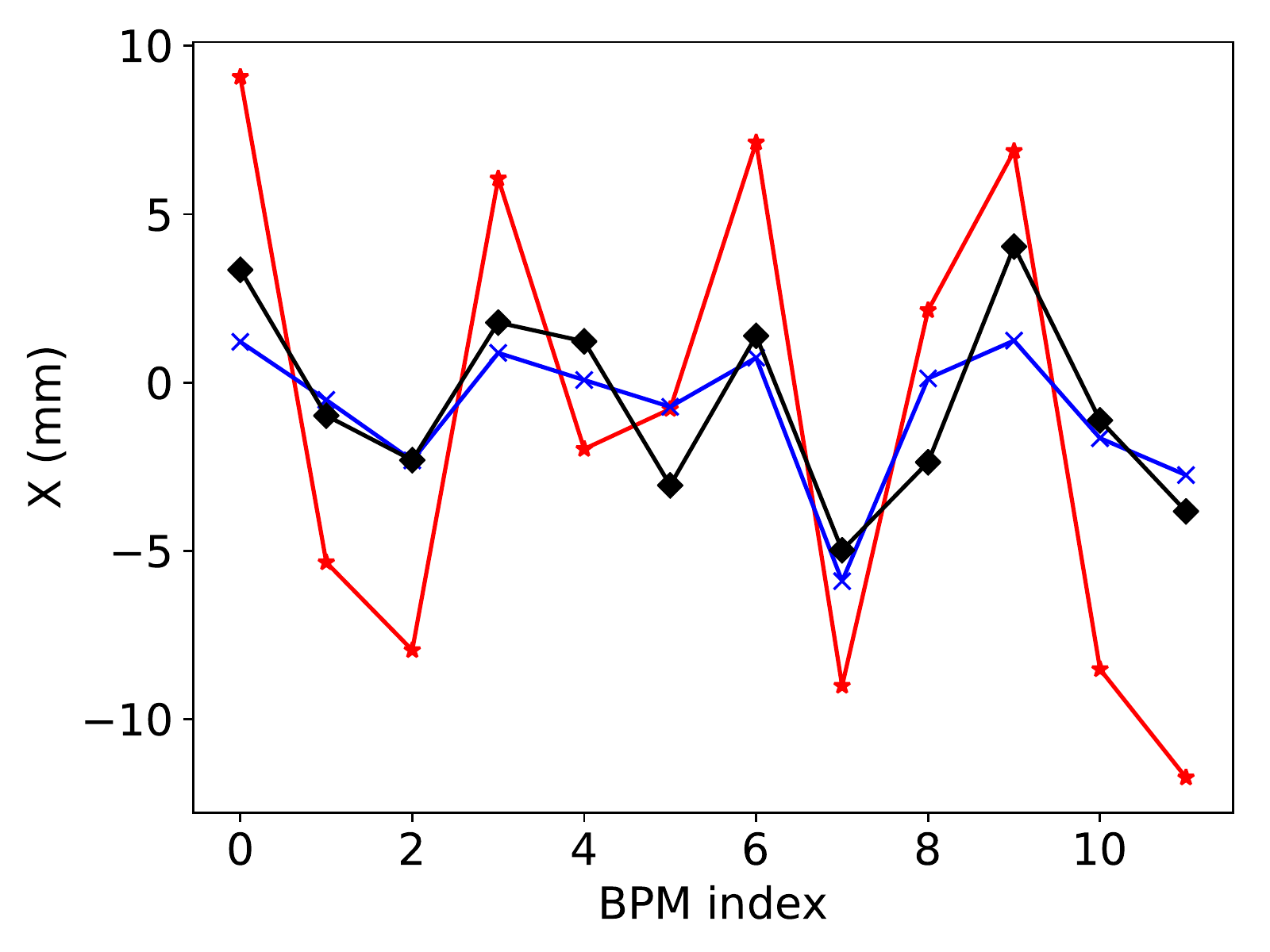}
\caption{Measured closed orbits in the $x$-plane of SIS18. Red(`*'): Perturbed orbit (RMS=7.12 mm). Blue(`x'): Corrected orbit using original ORM (RMS=2.15 mm). Black(`$\diamond$'): Corrected orbit using nearest-Circulant approximation (RMS=2.82 mm).\label{fig14_1}}
\end{figure}

In many scenarios, the Circulant symmetry of the ORM can be broken due to the odd placement of BPMs and correctors, presence of insertion devices, beta beating etc. For example, two horizontal correctors in SIS18 are placed in the second dipoles of the $4^{th}$ and $6^{th}$ sections while all others are in the first dipoles, hence breaking the Circulant symmetry in the corresponding columns highlighted with red color in the ORM below calculated by MAD-X~\cite{madx} for triplet optics in the units of $mm/mrad$:
\begin{equation}
    \setcounter{MaxMatrixCols}{20}
    \label{eq23a}
    \textbf{R}_{x}=\setlength\arraycolsep{1.5pt}
    \tiny
\begin{bmatrix}
-2.78&-4.86&-1.24&\hspace{2mm}\textcolor{red}{2.89}&-6.89&\hspace{2mm}\textcolor{red}{3.82}&\hspace{2mm}4.02&-7.36&\hspace{2mm}5.27&\hspace{2mm}0.70&-6.16&\hspace{2mm}7.07\\
\hspace{2mm}7.07&-2.77&-4.87&\hspace{2mm}\textcolor{red}{1.66}&\hspace{2mm}6.45&\textcolor{red}{-5.32}&\hspace{2mm}2.27&  \hspace{2mm}4.04&-7.36&\hspace{2mm}5.26&\hspace{2mm}0.71&-6.16\\
-6.15&\hspace{2mm}7.06&-2.77&\textcolor{red}{-4.98}&-1.25&\hspace{2mm}\textcolor{red}{2.9}&-6.88&\hspace{2mm}2.26&\hspace{2mm}4.03&-7.35&\hspace{2mm}5.26&\hspace{2mm}0.71\\
\hspace{2mm}0.70&-6.15&\hspace{2mm}7.08&\textcolor{red}{-4.08}&-4.87&\hspace{2mm}\textcolor{red}{1.65}&\hspace{2mm}6.44&-6.89&\hspace{2mm}2.27&\hspace{2mm}4.02&-7.36&\hspace{2mm}5.27\\
\hspace{2mm}5.26& \hspace{2mm}0.71&-6.16&\hspace{2mm}\textcolor{red}{5.26}&-2.77&\textcolor{red}{-4.99}&-1.25&\hspace{2mm}6.45&-6.89& \hspace{2mm}2.26&\hspace{2mm}4.04&-7.36\\
-7.35&\hspace{2mm}5.26&\hspace{2mm}0.71&\textcolor{red}{-2.56}&\hspace{2mm}7.07&\textcolor{red}{-4.08}&-4.86&-1.25&\hspace{2mm}6.44&-6.88&\hspace{2mm}2.26&\hspace{2mm}4.03\\
\hspace{2mm}4.02&-7.36&\hspace{2mm}5.27&\textcolor{red}{-2.03}&-6.16&\hspace{2mm}\textcolor{red}{5.26}&-2.78&-4.87&-1.24&\hspace{2mm}6.44&-6.89&\hspace{2mm}2.27\\
\hspace{2mm}2.26&\hspace{2mm}4.03&-7.36&\hspace{2mm}\textcolor{red}{5.12}&\hspace{2mm}0.71&\textcolor{red}{-2.57}&\hspace{2mm}7.07&-2.77&-4.87&-1.25&\hspace{2mm}6.45&-6.90 \\
-6.89&\hspace{2mm}2.26&\hspace{2mm}4.03&\textcolor{red}{-4.44}&\hspace{2mm}5.26&\textcolor{red}{-2.02}&-6.15&\hspace{2mm}7.07&-2.78&-4.86&-1.25&\hspace{2mm}6.44\\
\hspace{2mm}6.44&-6.89&\hspace{2mm}2.27&\hspace{2mm}\textcolor{red}{0.49}&-7.36&\hspace{2mm}\textcolor{red}{5.12}&\hspace{2mm}0.70&-6.16&\hspace{2mm}7.08&-2.78&-4.87&-1.24\\
-1.24&\hspace{2mm}6.44&-6.9&\hspace{2mm}\textcolor{red}{3.82}&\hspace{2mm}4.04&\textcolor{red}{-4.45}&\hspace{2mm}5.26&\hspace{2mm}0.71&-6.16&\hspace{2mm}7.07&-2.77&-4.87\\
-4.86&-1.25&\hspace{2mm}6.44&\textcolor{red}{-5.31}&\hspace{2mm}2.26&\hspace{2mm}\textcolor{red}{0.50}&-7.35&\hspace{2mm}5.26&\hspace{2mm}0.71&-6.15&\hspace{2mm}7.06&-2.77\\
\end{bmatrix}
\end{equation}
In this case, a DFT of only one row or column cannot be used directly for the decomposition and inversion of the ORM. However, a slight modification of the ORM in order to find a nearest-Circulant approximation is proposed as an alternate for broken symmetries. This is based upon the fact an iterative correction implemented for most orbit correction systems can still converge with a modified process model at the cost of more iterations or correction speed~\cite{ipac18, nima}. A recent example being the use of Tikhonov regularization~\cite{TK1} in order to improve the robustness and stability margins of the controller by modifying the gain for higher order modes of ORM by factors $p_{i}$ given as~\cite{sandira2} 
    \begin{equation}\label{eq30}
        \frac{p_{i}}{s_{i}}=\frac{1}{s_{i}}\cdot\frac{s_{i}^{2}}{s_{i}^{2}+\mu}
    \end{equation}
where $s_{i}$ are the singular values and an appropriate value of $\mu$ $>0$ serves to decrease the condition number of the ORM defined as the ratio of its largest to smallest singular values~\cite{condnum}. \\
The theory of nearest-Circulant approximation is discussed in detail in~\cite{nc} but has never been explored for ORM inversion before. For a given square matrix \textbf{R}, its nearest-Circulant approximation $\textbf{R}_{NC}$ = $circ[c_{0},c_{1},...c_{n-1}]$ can be found by the Frobenius inner product of \textbf{R} with permutation matrices $\pi_n^{i}$ as
\begin{equation}
    \label{eq24}
    \centering
    c_i = \frac{1}{n}<\textbf{R},\pi_n^{i}>
\end{equation}
where $n$ is the size of the matrix and the order of the permutation matrix is $i = 0,...,n-1$. Eq.~\ref{eq24} is equivalent to the averaging of the diagonal elements of \textbf{R} and for the resultant approximation, the theory discussed in section~\ref{seccrc} holds. Figure~\ref{fig14} shows the singular values of the SIS18 horizontal ORM and its nearest-Circulant approximation for a qualitative comparison.
The effect of the nearest-Circulant approximation seen as a "model deviation" on the orbit correction can be quantified in terms of the residual after one iteration of orbit correction, given as~\cite{svd1,ipac18} 
\begin{equation}\label{eq24a}
    \centering
    r_\text{1} =  z - \textbf{R}_{M}\Theta_{NC} = z - \textbf{R}_{M} \textbf{R}_{NC}^{-1}~z = \left(\textbf{I}-\textbf{R}_{M}\textbf{R}_{NC}^{-1}\right) z
\end{equation}
where $\textbf{R}_{M}$ is the actual machine ORM and $\Theta_{NC}$ is the corrector settings vector calculated using the nearest-Circulant approximation $\textbf{R}_{NC}$ of the original model ORM $\textbf{R}$. Figure~\ref{fig14_1} shows the experimentally measured closed orbit before and after one iteration of correction in the horizontal plane of the SIS18. The difference in the RMS of the residual orbit is $\simeq 9 \%$ of the initial distortion for using the original ORM and its nearest-Circulant approximation. It is likely that a controller which is capable of orbit correction for the original ORM will also provide the correction for its nearest-Circulant approximation. 
\subsection{Block Circulant symmetry of SIS100 ORMs}
\begin{figure}[h!]
\centering
\includegraphics[width=8.6 cm]{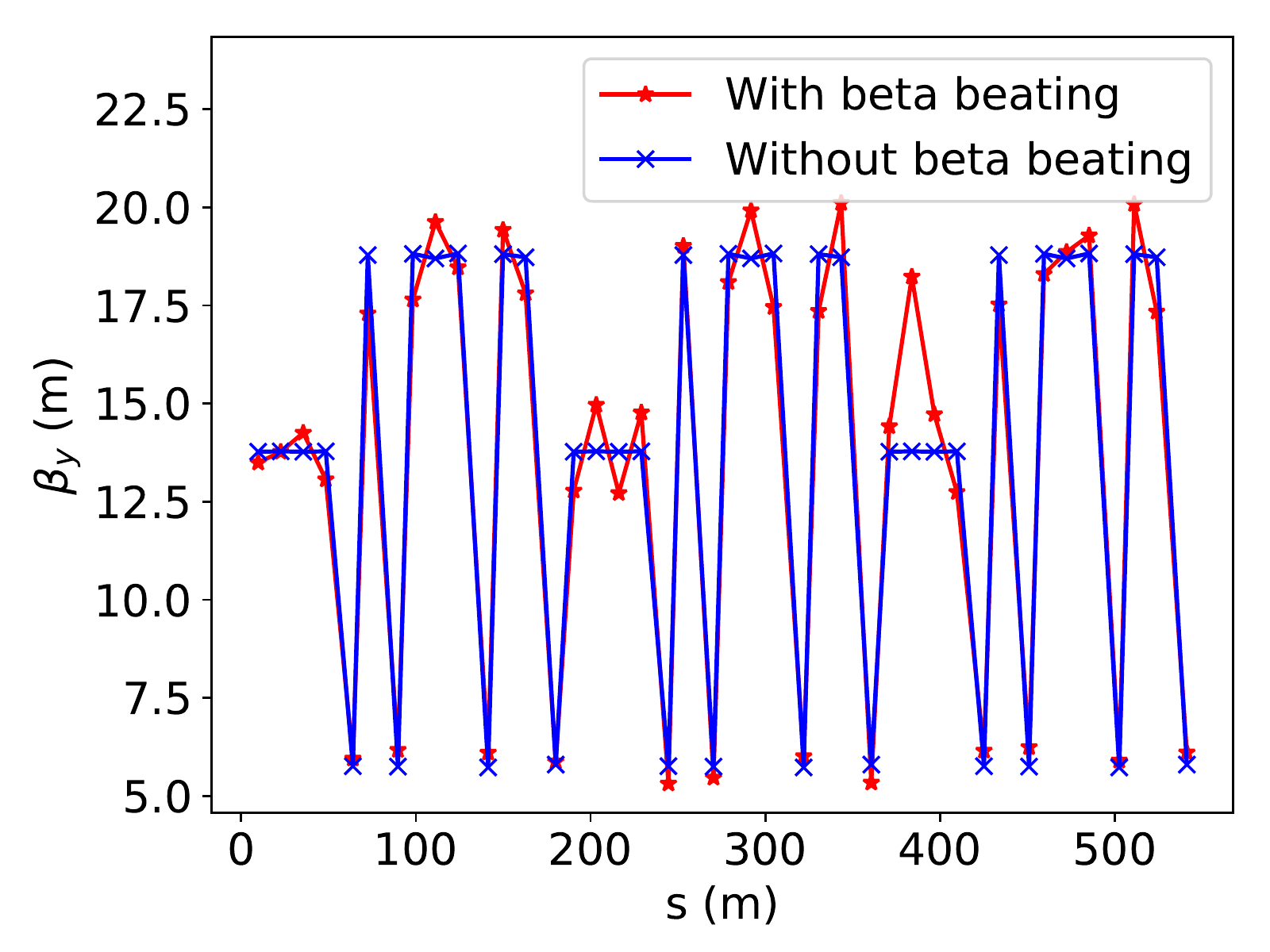}
\caption{Beta function at BPM locations in the y-plane of SIS100 (simulated in MAD-X) plotted for three cells, with beta beating (red(`*')) and without beta beating (blue(`x')). The abscissa is the distance along the synchrotron.\label{fig4}}
\end{figure}

\begin{figure}[h!]
\centering
\includegraphics[width=8.6 cm]{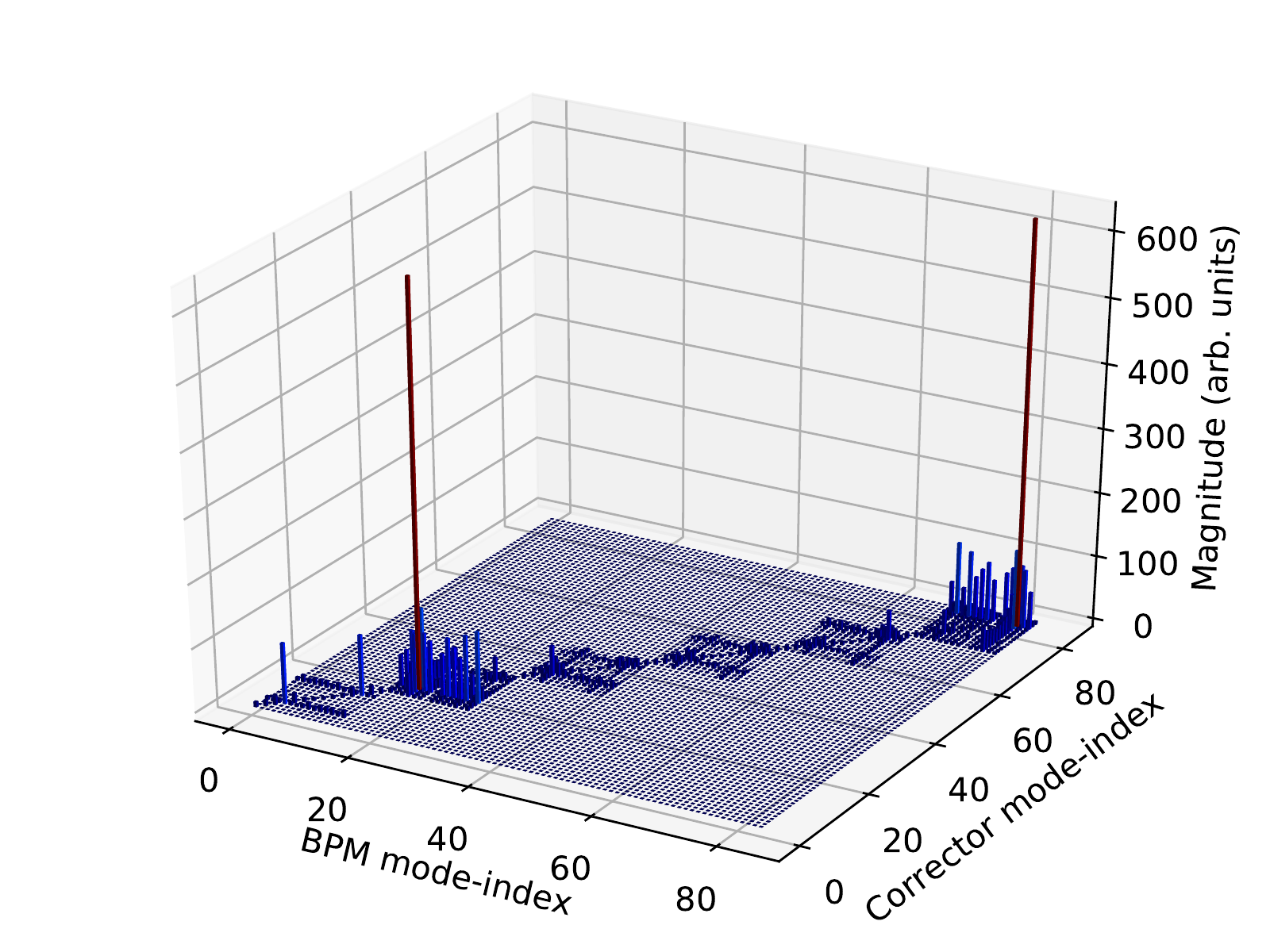}
\caption{Magnitudes of the complex entries of the block diagonal matrix $\textbf{D}$ defined in Eq.~\ref{eq15_1}, calculated for the case of ignoring the beta beating in the $x$-plane of SIS100.\label{fig7}}
\end{figure}

\begin{figure}[h!]
\centering
\includegraphics[width=8.6 cm]{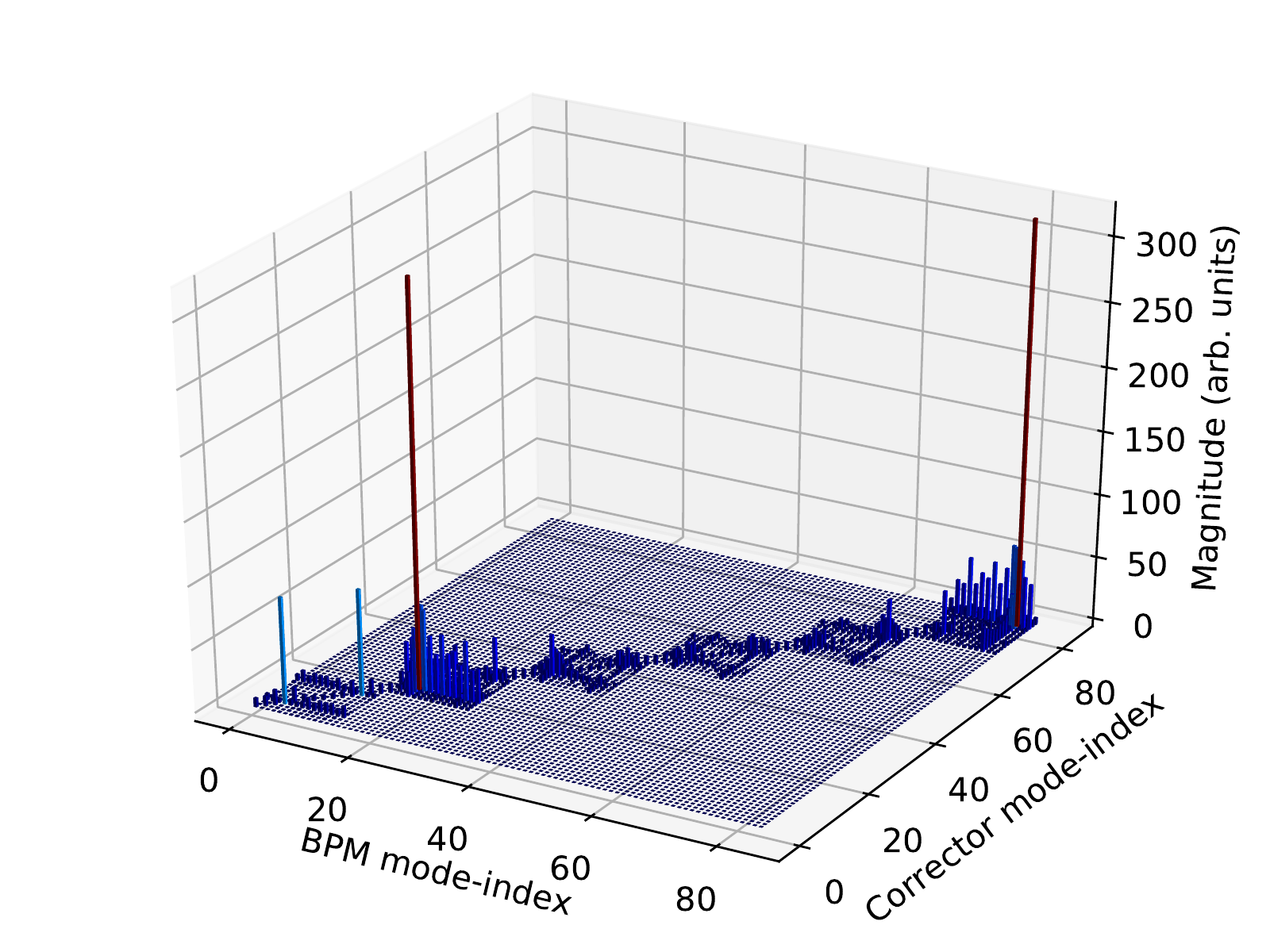}
\caption{Magnitudes of the complex entries of the block diagonal matrix $\textbf{D}$ defined in Eq.~\ref{eq15_1}, calculated for the case of ignoring the beta beating in the $y$-plane of SIS100.\label{fig8}}
\end{figure}

\begin{figure}[h!]
\centering
\includegraphics[width=8.6 cm]{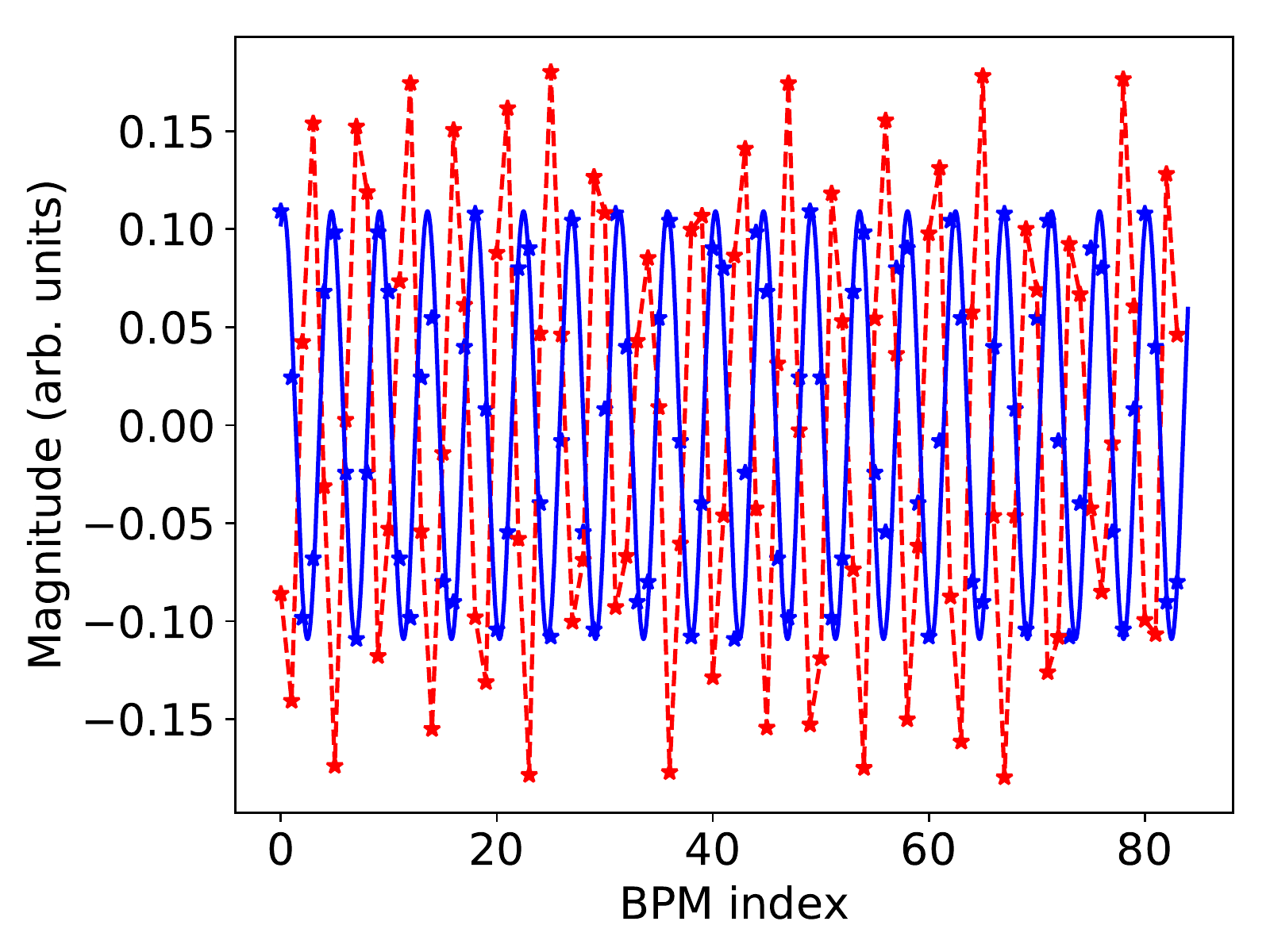}
\caption{Tune modes of DFT matrix ($\textbf{F}_{6}^{*} \otimes \textbf{F}_{14}^{*})$ (blue(solid)) and SVD matrix \textbf{U} (red(dotted)) for SIS100 ORM in the x-plane. The horizontal tune is 18.87.\label{fig11}}
\end{figure}

\begin{figure}[h!]
\centering
\includegraphics[width=8.6 cm]{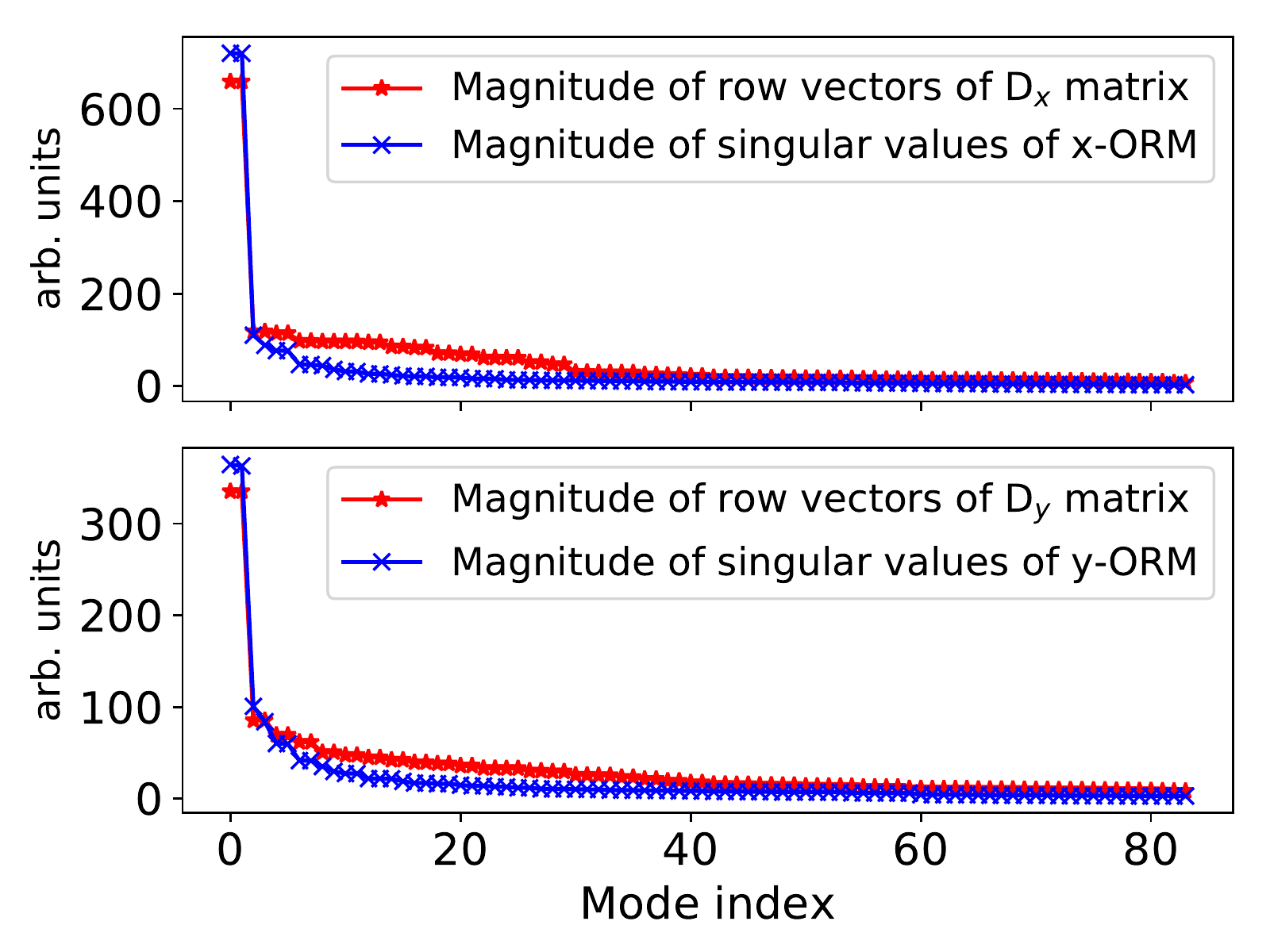}
\caption{Comparison of the SVD singular values with the magnitudes of the rows of matrix $\textbf{D}$ defined in Eq.~\ref{eq15_1} for the SIS100 ORM in x-plane. The abscissa is the index of singular values of the matrix $\textbf{S}$ as well as the index of rows of matrix $\textbf{D}$ (after sorting in the descending order of their magnitude).\label{fig12}}
\end{figure}

The SIS100 is the largest synchrotron of the FAIR project with a six-fold symmetry. Each of the six sections has 14 BPMs and 14 correctors while in one section, the cold quadrupole is replaced by a warm quadrupole~\cite{warm}. The warm quadrupole results in a beta beating (peak-peak $\approx 10 \%$) and hence a loss of symmetry in the ORM as calculated by MAD-X~\cite{madx}. The beta function at BPM locations in $y$-plane has been plotted in Fig.~\ref{fig4} for three consecutive sections with and without beta beating (by replacing the warm quadrupole by a cold quadrupole in MAD-X~\cite{madx}). The block symmetry of the SIS100 ORMs can be explored in two ways; either by a) ignoring beta beating or by b) finding the nearest-block Circulant approximation by averaging the diagonal blocks.
In either case, the ORM attains a block-wise symmetry such that identical blocks of elements appear at the diagonal locations:
\begin{equation}
    \label{eq14}
    \textbf{R}_{BC}=circ[\textbf{A}_0,\textbf{A}_1,\textbf{A}_2.......\textbf{A}_{m-1}]
\end{equation}
 Such a matrix is referred to as block Circulant matrix (BCM)~\cite{circulant}. Here $\textbf{A}_i$ are the arbitrary square blocks of dimension $n$ and $\textbf{R}_{BC}$ is a square matrix of dimensions $mn\times mn$. For the SIS100 ORM, $m$ is the number of sections while $n$ is the number of BPMs or correctors in each section. Here, we shall limit the discussion to the square BCM resulting from the equal number of BPMs and correctors in each section. Many machines have unequal number of BPMs and corrector per section, which can also be worked out, and a specific example of $n_{bpm} =  2n_{corrector}$ is shown in Appendix C. 
A square BCM can also be diagonalized with the help of Fourier matrices as~\cite{circulant}
\begin{equation}
    \label{eq15}
    \centering
    \textbf{R}_{BC}=(\textbf{F}_{m} \otimes \textbf{F}_{n})^{*}\textbf{D}(\textbf{F}_m \otimes \textbf{F}_n)
\end{equation} 
where
\begin{equation}
    \label{eq15_1}
    \centering
    \textbf{D}=\begin{bmatrix}
        \textbf{M}_0&0&0&.&.&0 \\
        0&\textbf{M}_1&0&.&.&0\\
        .&.&.&.&.&0 \\
        .&.&.&.&.&0 \\
        0&0&0&.&.&\textbf{M}_{m-1} \\
\end{bmatrix}
\end{equation}
$\textbf{F}_m$ and $\textbf{F}_n$ are the standard Fourier matrices defined in Eq.~\ref{eq7}. The symbol $\otimes$ denotes the Kronecker product of the matrices. $\textbf{M}_i$ are the square matrices of dimension $n$ which contain all the information of the block Circulant matrix and can be calculated using only the first row of blocks as reproduced from~\cite{circulant} in appendix B.
Equation~\ref{eq15} can be solved to calculate the inverse or pseudo-inverse (\textbf{R}$^{+}_{BC}$) of the ORM as
\begin{equation}
    \label{eq22}
    \centering
    \textbf{R}^{+}_{BC} = (\textbf{F}_{m} \otimes \textbf{F}_{n})^{*}(\textbf{D}^{+})(\textbf{F}_{m} \otimes \textbf{F}_{n}) 
\end{equation}
where
\begin{equation}
    \label{eq23}
    \centering
    \textbf{D}^{+} = diag(\textbf{M}_0^{+}, \textbf{M}_1^{+},....,\textbf{M}_{m-1}^{+})
\end{equation}
The block diagonal matrices ($\textbf{D}$) of SIS100 ORMs for both the planes in case of ignoring the beta beating are visualized in Fig.~\ref{fig7} and ~\ref{fig8} as Lego plots of their absolute values, respectively. It can be seen that very few elements have larger magnitude providing the liberty to remove weak modes before inversion just like singular values in SVD. Also the rows of the block diagonal matrices containing the largest absolute values correspond to the tune modes of the Fourier matrices ($\textbf{F}_{m} \otimes \textbf{F}_{n}$) which are plotted for both planes in Fig.~\ref{fig11} along with tune modes of SVD. Moreover, a comparison of the singular values with the moduli of the row vectors of the block diagonal matrices in Fig.~\ref{fig12} provides a clear hint on the correspondence between the two methods.
The deviation of the block Circulant approximations of the SIS100 ORMs from the original matrices can also be characterized using Eq.~\ref{eq24a}. The first iteration residuals (\%) for the nearest-block Circulant approximation, the block Circulant approximation by ignoring the beta beating and the nearest-Circulant approximation calculated by Eq.~\ref{eq24} (for comparison) have been listed in table~\ref{tab1}. One can see that the block Circulant approximation by either of aforementioned techniques can be used for the orbit correction.
\begin{table}[h!] 
 \caption{Calculated first iteration residuals in the case of symmetry exploitation in the SIS100 ORMs\label{tab1}}
 \begin{tabular}{l c c c}\hline\hline
   SIS100 ORM     & $x$-plane & $y$-plane \\ \hline
    Original ORM    & 0 $\%$ &  0 $\%$\\\hline
    Nearest-Circulant approximation & 34$\%$ & 31 $\%$  \\  \hline
Nearest-block Circulant approximation&13$\%$& 17 $\%$\\ \hline
Ignoring beta-beating&16$\%$& 20 $\%$\\ \hline
\end{tabular} \\
 \end{table}
\section{Applications and discussion}
Symmetry exploitation in the ORM for the DFT-based decomposition has some practical advantages over SVD-based decomposition. In this section, the benefits concerning computational complexity, robustness against ``missing'' BPMs (e.g. ignoring a possibly faulty position reading), uncertainty modeling and removal of dispersion induced effect from the closed orbit are demonstrated with the help of simulations and experiments for the SIS18 synchrotron of GSI. 
\subsection{Computational complexity}
DFT-based diagonalization and inversion of the ORM has significant computational benefit over SVD. For a square matrix of dimension $n$, the numerical complexities for the SVD of the whole matrix and the DFT of its first row are $\mathcal{O}(4 n^{3})$~\cite{svdtime} and $\mathcal{O}(n^{2})$~\cite{svd}, respectively. Such a reduction in numerical complexity becomes meaningful for the larger ORMs and for the on-ramp orbit correction. SIS18 is an example~\cite{ipac18} of the later, where lattice settings as well as ORM change systematically during the ramp. For SVD, one has to update all three matrices \textbf{U}, \textbf{V} and \textbf{S} for the ORM variation while for DFT, it is a set of only $n$ numbers to be updated in the diagonal matrix. This could lead to significant memory usage reduction if the matrix multiplication for a continuously changing matrices is implemented in an FPGA.
\subsection{Orbit correction in the case of malfunctioned BPMs}
\begin{figure}[h!]
\centering
\includegraphics[width=8.6 cm]{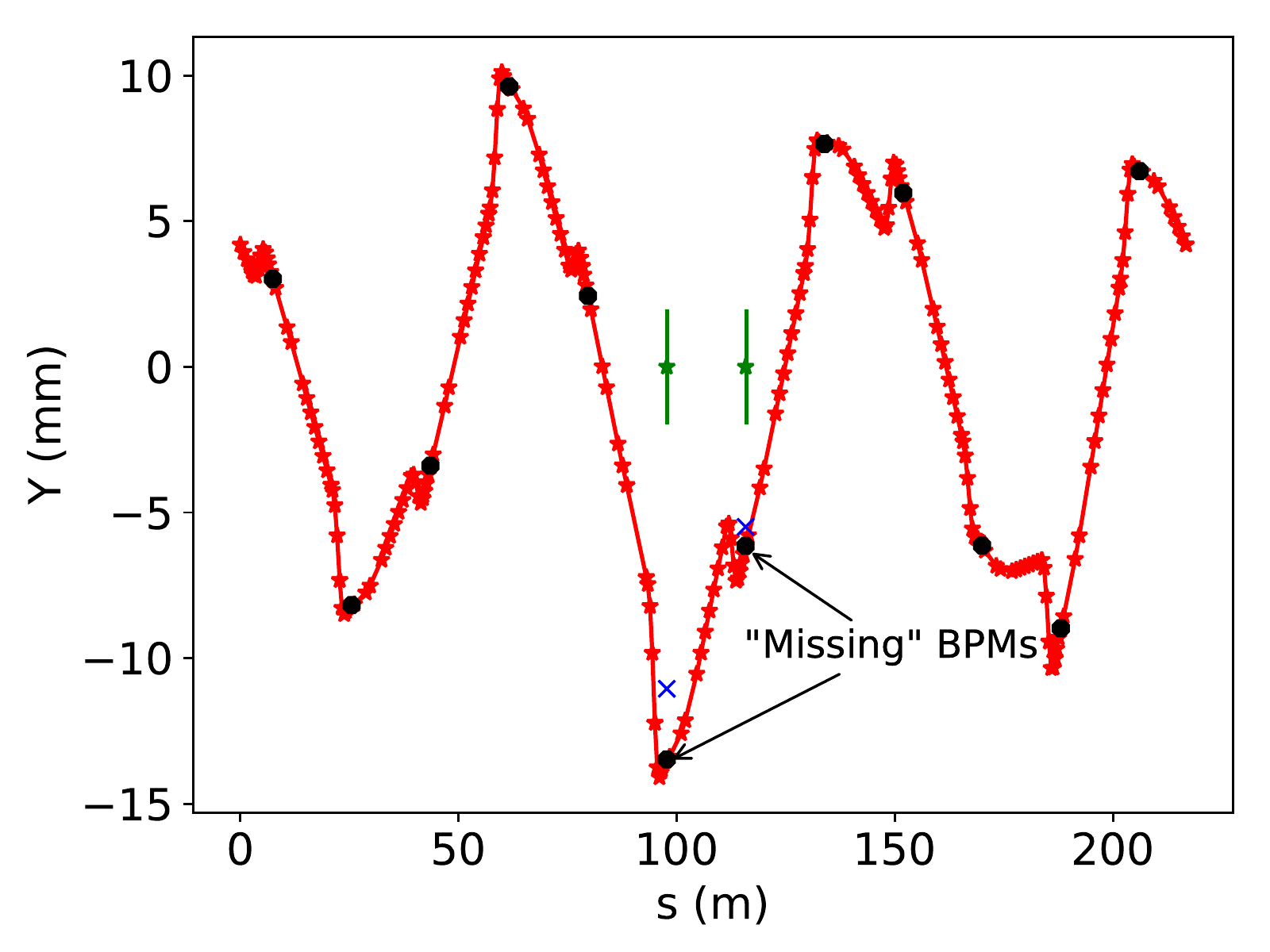}
\caption{Prediction of the closed orbit position at the ``missing'' BPM locations. Red(`*'): Simulated perturbed orbit in MAD-X in the y-plane of SIS18. Black(`o'): Sampling of perturbed orbit at BPM locations. Green(`*'): Random initial guess for the orbit position at the two ``missing'' BPM locations. Blue(`x'): Estimated orbit position at ``missing'' BPM locations using DFT mode structure of the ORM.\label{fig16}}
\end{figure}

The physical interpretation of BPM and corrector mode space provided by DFT-based decomposition of ORM can be used to interpolate the closed orbit at the location of some malfunctioning or ``missing'' BPM. This is demonstrated for the SIS18 in vertical plane using MAD-X for a scenario of two consecutive BPMs being excluded. The operational scenario of BPM electronics failure due to radiation shower happens often in hadron synchrotrons and can lead to local bumps if there is not enough redundancy in the number of BPMs.\\
Figure \ref{fig16} shows a simulated closed orbit (red curve) perturbed as a result of random vertical misalignments (in a range of -0.85 to 0.85 mm) in all 24 quadrupoles of the SIS18. A misaligned quadrupole has the effect of a magnetic dipole on the beam and as a result the closed orbit is perturbed from its ``ideal path''. 
 \begin{figure}[h!]
\centering
\includegraphics[width=8.6 cm]{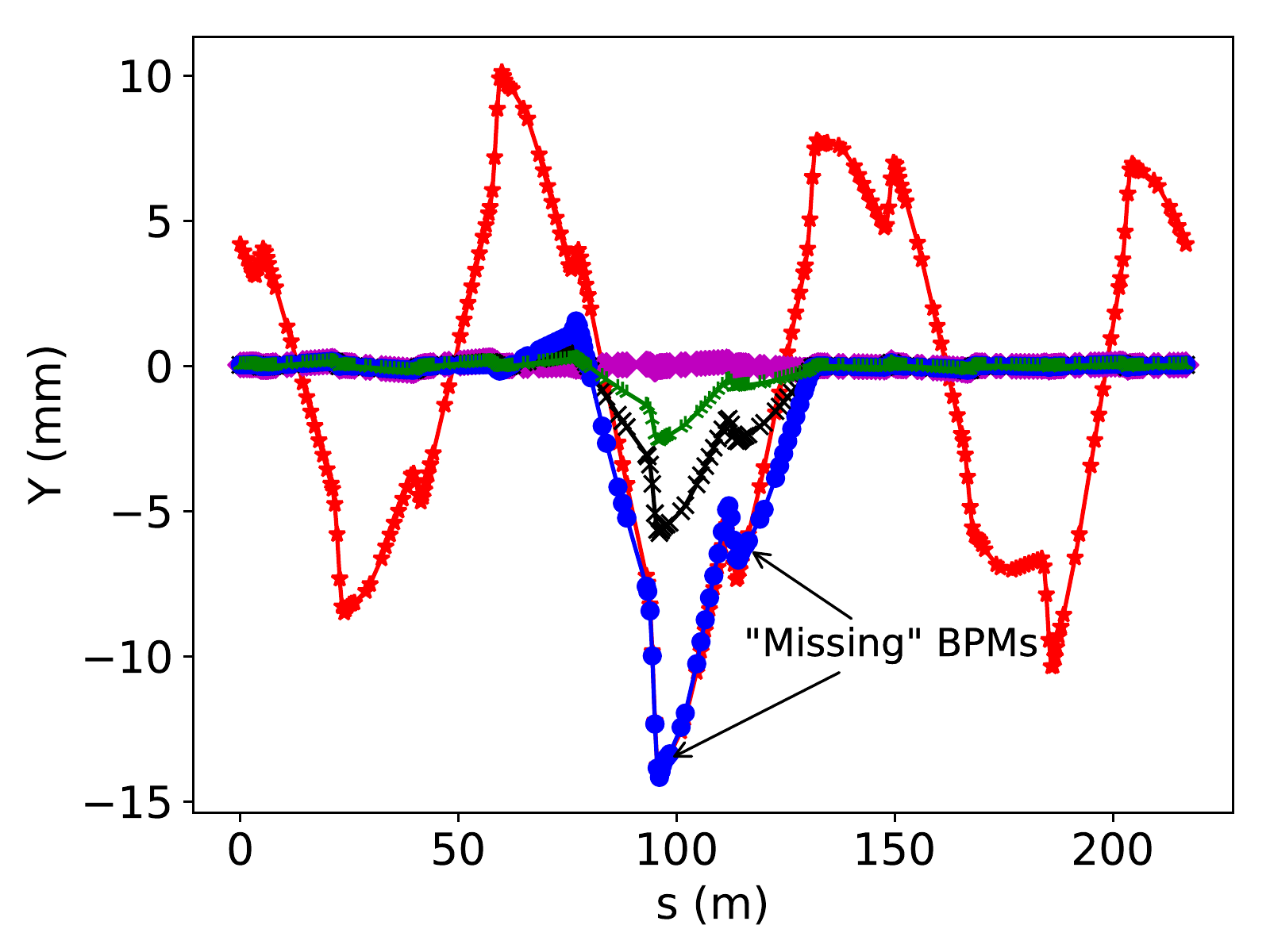}
\caption{Simulation of orbit correction for various cases of ``missing'' BPMs. Red(`*'): Simulated perturbed orbit in MAD-X in y-plane of SIS18. Magenta(`$\diamond$'): Corrected orbit using all BPMs. Green(`$\Yup$'): Corrected orbit using the predicted orbit position for the ``missing'' BPMs. Black(`x'): Corrected orbit excluding the rows corresponding to the ``missing'' BPMs from ORM. Blue(`o'): Corrected orbit by taking zero orbit position at the ``missing'' BPM locations.\label{fig17}}
\end{figure}
The black dots in Fig.~\ref{fig16} represent the sampling of the perturbed orbit at the locations of BPMs while the green dots show the mean values of random orbit positions (in a range of -1 to 1 mm marked as error bars around the green dots) assumed at the location of ``missing'' or faulty BPMs. A combination of cosine functions with discrete frequencies $f=3,4,2,5$ (corresponding to dominant Fourier modes) 
was used to estimate the orbit position at the ``missing'' BPM locations keeping their relative amplitudes and phases as free parameters to be optimized. The fitting algorithm was constrained to keep the fitted curve closest to the orbit positions within an accuracy of 0.01 mm at the working BPM locations while free to choose any value at the location of ``missing'' BPMs. As a result, the optimized orbit positions are found to be closer to the actual orbit positions within maximum errors of 3 $\pm$ 0.048 mm and 0.5 $\pm$ 0.048 mm, respectively.
Figure \ref{fig17} shows the simulated orbit correction using the estimated orbit positions (green curve). The red curve shows the perturbed orbit without corrections while the magenta curve shows the corrected orbit when all BPMs are working. The orbit correction using SVD of the non-Circulant matrix (excluding the rows corresponding to the faulty BPMs from ORM) is plotted in black color. The orbit correction taking the orbit position ``zero'' at the ``missing'' BPM locations and using a Circulant matrix is also plotted as a blue curve for comparison. The robustness against ``missing'' BPMs is shown by the overall improved correction obtained using an estimated beam position instead of using the non-Circulant matrix. Besides the better global correction one can also get the benefits of Circulant symmetry and DFT-based decomposition (e.g. online decomposition during ramp) even when the symmetry has been broken due to the ``missing'' BPMs.
\subsection{Uncertainty description in spatial process model}
Uncertainties appear into ORMs through various sources e.g. BPM and corrector calibration errors, tune variation due to magnet gradient errors or during acceleration ramp~\cite{ipac18}. For SVD, the uncertainty in the ORM appears in all the three matrices of SVD represented by $\Delta$~\cite{sandira1}
\begin{equation}
    \label{eq27}
    \centering
    (\textbf{I}+\Delta_{R})\textbf{R}= (\textbf{I}+\Delta_{U})\textbf{U}(\textbf{I}+\Delta_{S})\textbf{S}\textbf{V}^{\text{T}}(\textbf{I}+\Delta_{V})^{\text{T}}
\end{equation}
Alternatively, Fourier coefficients of harmonic analysis given as~\cite{sandira2} 
    \begin{equation}
    \label{eq29}
    \sigma_f = \frac{Q_{z}}{\pi (Q_{z}^2 - f^2)}
    \end{equation}
where $f$ is the discrete frequency and $Q_{z}$ is the betatron tune in either transverse plane, have also been used to express uncertainty in the betatron tune. But this method cannot be accompanied with matrix inversion and Fourier coefficients given in Eq.~\ref{eq29} have no quantitative relation with SVD singular values.
The information compression into one diagonal matrix in case of DFT-based diagonalization, can provide a significant simplification of the uncertainty description as 
\begin{eqnarray}
    \label{eq28}
    \centering
    (\textbf{I}+\Delta_{R_{C}})\textbf{R}_{C}&=& \textbf{F}^{*}(\textbf{I}+\Delta_{\mathbf{\Lambda}})\mathbf{\Lambda}\textbf{F}\\
    (\textbf{I}+\Delta_{R_{BC}})\textbf{R}_{BC}&=& (\textbf{F}_{m}^{*} \otimes \textbf{F}_{n}^{*})(\textbf{I}+\Delta_{D})\textbf{D}(\textbf{F}_{m} \otimes \textbf{F}_{n})
\end{eqnarray}
\subsection{Momentum mismatch and orbit correction}
\begin{figure}[h!]
\centering
\includegraphics[width=8.6 cm]{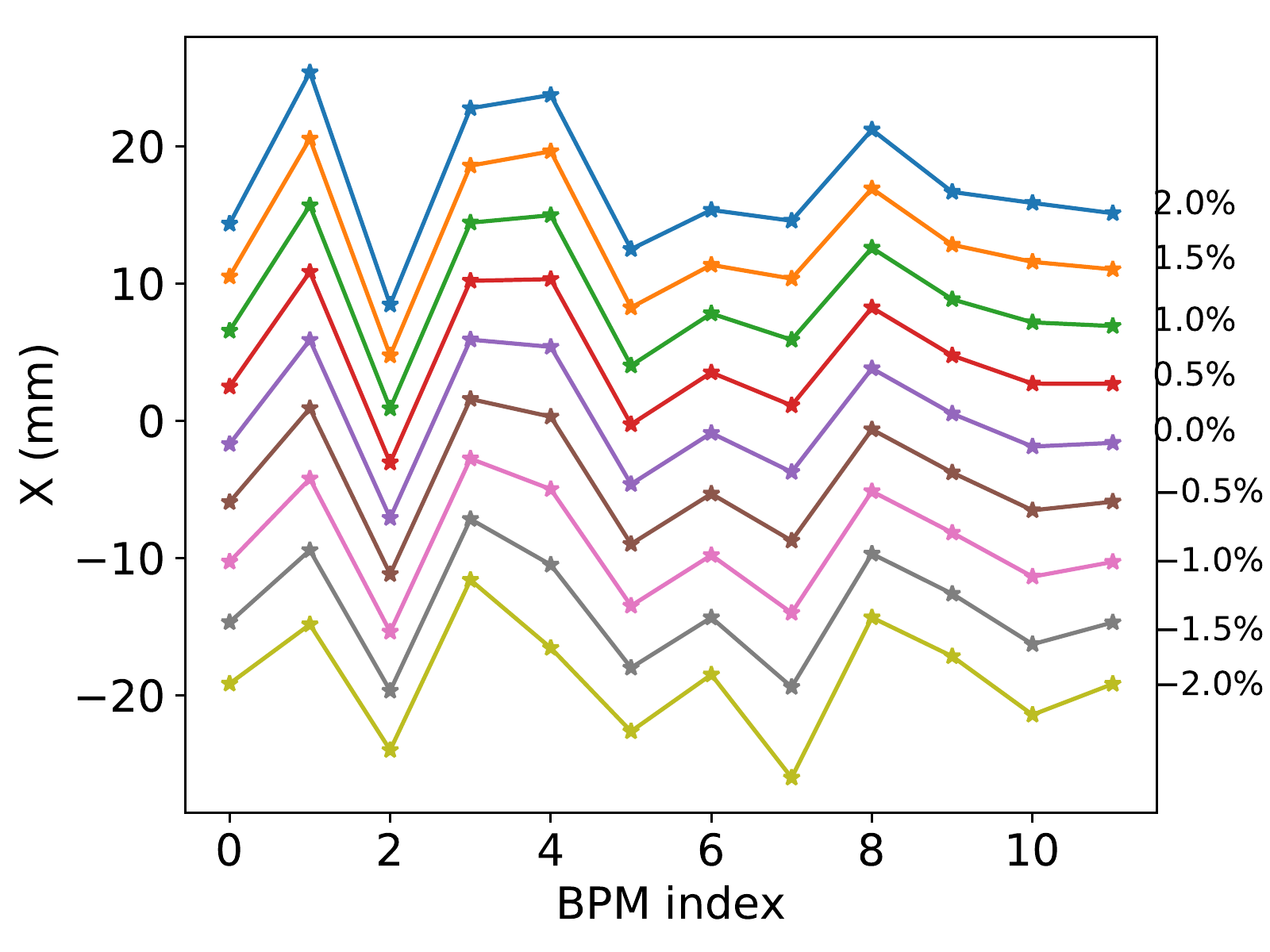}
\caption{Measured horizontal closed orbits in SIS18 for the various average momentum offsets represented as $\%$ with each orbit.\label{fig18}}
\end{figure}

\begin{figure}[h!]
\includegraphics[width=8.6 cm]{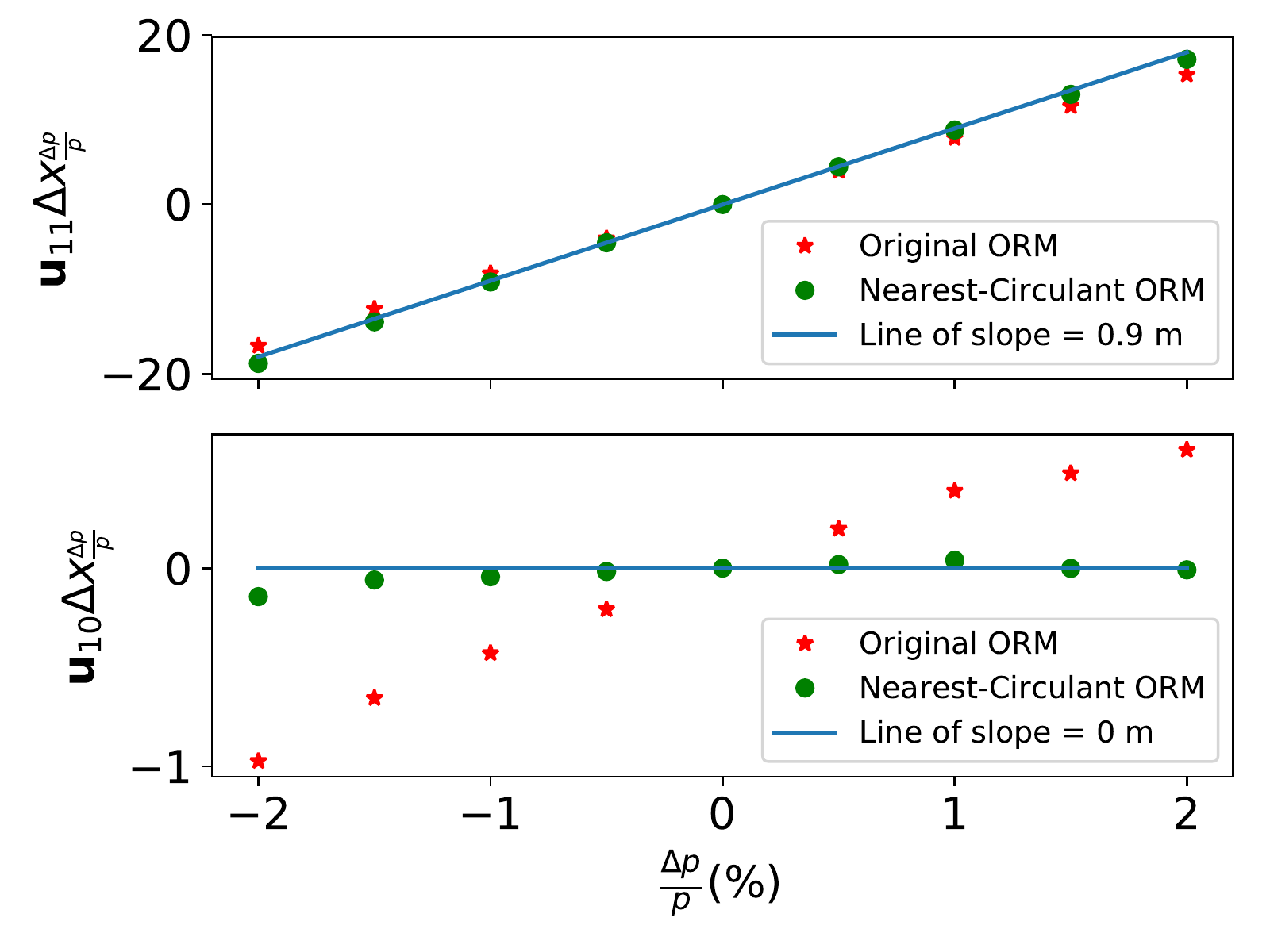}
\caption{Coupling of experimentally measured DC shift $\Delta x$ (mm) in the horizontal closed orbit to the last two columns of the \textbf{U} matrices ($\textbf{u}_{11}$ and $\textbf{u}_{10}$) of the original ORM and its nearest-Circulant approximation.\label{fig19}}
\end{figure}

A mismatch between RF frequency and the dipole field results in a relative mismatch in the average momentum of the beam $\frac{\Delta p}{p}$ and the closed orbit deviates from the equilibrium position primarily in the x-plane (in the absence of coupling between the two planes)~\cite{dispbk}:
\begin{equation}
    \label{eq31}
    \centering
    \Delta x_{D}(s) = D_{x}(s)\frac{\Delta p}{p}
\end{equation}
Here $D_{x}(s)$ is the dispersion function and $\Delta x_{D}(s)$ is the resultant DC (constant) shift in the closed orbit. Fig.~\ref{fig18} shows a set of measured horizontal closed orbits for the induced relative momentum mismatch in the range of -2$\%$ to 2$\%$ with a step size of 0.5$\%$ in SIS18 at extraction energy. One can see a shift of the mean value of the closed orbit as a function of $\frac{\Delta p}{p}$. An attempt to correct such an orbit shift can saturate the corrector magnets. Therefore, the contribution of such a displacement needs to be subtracted from the closed orbit before correction. Generally, the dispersion induced orbit distortion will couple into many modes of BPM space, depending on the sampling of the dispersion function. For the SIS18, the dispersion function has the same value $D(s)=D_{0}$ at all the BPM locations and the resultant dispersion induced DC part of the closed orbit will couple to the pure DC mode of the BPM space corresponding to $f=0$ in case of Circulant ORM and can be ignored by removing the singular value corresponding to that mode (the last column of $\textbf{U}$). Fig.~\ref{fig19} shows the coupling of the DC part of the measured closed orbits to the last two modes of the original non-Circulant ORM and its nearest-Circulant approximation for SIS18. It is evident that DC shift of the closed orbit couples with the last mode of the matrix $\textbf{U}$ as $\textbf{u}_{11}\Delta x_{\frac{\Delta p}{p}}$ increases linearly with $\frac{\Delta p}{p}$ with a slope of 0.9 m (equal to the value of dispersion function $D_0$ at the BPM locations). The slight deviation from this behavior in case of original ORM is due to a slight distortion of last mode from pure DC structure due to the broken symmetry. This results in the coupling of the DC part of the closed orbit to higher modes as well, as evident in Fig.~\ref{fig19}. Here in case of nearest-Circulant approximation there is almost zero coupling to the $2^{nd}$ last mode. This shows the benefit of Circulant symmetry in the ORMs in order to ignore the momentum mismatch-based DC shift of the closed orbit by simply discarding the last DC mode of the matrix \textbf{U}, instead of measuring and subtracting it from the closed orbit before correction. The obtained dispersion effect from the last mode can also be utilized for the synchronism of the rf frequency and the dipole magnetic field.
\section{Conclusion}
 An efficient method relying on the Circulant symmetry exploitation in the synchrotrons for the diagonalization and inversion of the orbit response matrix (ORM) is introduced in this paper. A Circulant ORM can be decomposed into BPM and corrector spaces with the help of a one dimensional DFT; this method is significantly faster than the singular value decomposition (SVD) and provides physical interpretation to the mode space. A relation between DFT coefficients and SVD singular values as well as modes is discussed which shows that DFT-based decomposition combines the benefits of harmonic analysis and SVD, in one method. Furthermore, the block Circulant symmetry arising from the block-wise periodicity of the lattice is also discussed primarily for SIS100 and the DFT-based decomposition for such ORMs is shown to have SVD-like robustness benefits in line with the pure Circulant symmetry case. These results are extended to broken symmetries with the nearest-Circulant approximation hence making it applicable to wider range of facilities. The reduced computational complexity of DFT can have benefits on the overall performance of closed orbit feedback systems in case of large ORMs and during the on-ramp orbit correction. The usefulness of the physical interpretation of BPM and corrector spaces as Fourier modes is demonstrated for the SIS18 synchrotron by interpolating the closed orbit at the location of ``missing'' BPMs and to achieve a better global orbit correction. DFT-based decomposition provides the information compression into the diagonal matrix only, which can significantly simplify the uncertainty modeling in the ORM, overcoming the major deficiency of SVD decomposition where the information is scattered in all three matrices. A new technique for the rejection of the dispersion induced shift of the closed orbit caused by the momentum mismatch during orbit correction is introduced and verified with the measurement at SIS18. Even though the ideas are presented in the context and with the aid of the FAIR synchrotrons, they would apply to most synchrotrons and storage rings with hints of aforementioned symmetry. Thus we would like to emphasize the importance of symmetry exploitation in the ORM and consideration at the design stage of the closed orbit feedback system.
 
 \section{Acknowledgments}
The authors would like to thank Thomas Reichert and Will Stem for reading the paper and Sajjad Mirza expresses gratitude to Guenther Rehm from Diamond Light Source UK, for useful discussions. This project has received funding from the European Union’s Horizon 2020 Research and Innovation programme under Grant Agreement No. 730871 (ARIES). Sajjad Mirza thankfully acknowledges the financial support from DAAD (Deutscher Akademischer Austauschdienst/German Academic Exchange Service) for his doctoral research work under Personal Reference No. $91605207$.
\appendix
\section{Permutation matrix}
A general permutation matrix $\pi_n^{i}$ is a square matrix of size $n$ which can be generated as a result of the cyclic permutations of the identity matrix. The number of permutations $i$ ($0 \leq i \leq n-1$) is called the order of the permutation matrix. Identity matrix is a permutation matrix of order $0$. Another example of permutation matrix of order 1 and size 4 can be written as  
\begin{equation}
    \label{eq5}
    \pi_4^{1}=\begin{bmatrix}
         0&1&0&0 \\ 
         0&0&1&0 \\ 
         0&0&0&1 \\ 
         1&0&0&0
\end{bmatrix}
\end{equation}
\section{Diagonalization of block Circulant matrices}
The following appendix is adopted from~\cite{circulant} (Theorem 56.4) as it is, where a BCM can also be written with the help of fundamental permutation matrices by replacing the numbers with matrices $\textbf{A}_i$ in Eq.~\ref{eq4} as
\begin{equation}
    \label{eq16}
    \centering
    \textbf{R}_{BC}=\sum_{i=0}^{n-1} (\pi_m^{i} \otimes \textbf{A}_{i}) 
\end{equation}
For arbitrary $\textbf{A}_i$
\begin{equation}
    \label{eq17}
    \centering
    \pi_m^{i} \otimes \textbf{A}_{i} = (\textbf{F}_{m}^{*}\Omega^{i}\textbf{F}_{m}) \otimes \textbf{F}_{n}^{*} (\textbf{F}_{n}\textbf{A}_{i}\textbf{F}_{n}^{*})\textbf{F}_{n}
\end{equation}
Let $\textbf{B}_i$ = $\textbf{F}_{n}\textbf{A}_{i}\textbf{F}_{n}^{*}$, the Eq.~\ref{eq16} becomes
\begin{equation}
    \label{eq18}
    \centering
    \textbf{R}_{BC} = (\textbf{F}_{m} \otimes \textbf{F}_{n})^{*}(\sum_{i=0}^{n-1} \Omega^{i} \otimes \textbf{B}_i)(\textbf{F}_{m} \otimes \textbf{F}_{n})  
\end{equation}
The middle term in Eq.~\ref{eq18} has a form of a diagonal matrix as
\begin{equation}
    \label{eq19}
    \centering
    \sum_{i=0}^{n-1} \Omega^{i} \otimes \textbf{B}_i = \text{diag}(\textbf{M}_{0}, \textbf{M}_{1}, ...., \textbf{M}_{m-1}) = \textbf{D}
\end{equation}
where $\textbf{M}_i$ can be calculated by the relation 
\begin{equation}
    \label{eq20}
    \begin{bmatrix}
        \textbf{M}_{0} \\
        \textbf{M}_{1}\\
        . \\ 
        . \\
         \textbf{M}_{m-1} \\
\end{bmatrix}=
(m^{\frac{1}{2}}\textbf{F}_{m}^{*} \otimes \textbf{I}_n)
\begin{bmatrix}
        \textbf{B}_{0} \\
        \textbf{B}_{1}\\
        . \\ 
        . \\
         \textbf{B}_{m-1} \\
\end{bmatrix}
\end{equation}
and Eq.~\ref{eq18} becomes 
\begin{equation}
    \label{eq21}
    \centering
    \textbf{R}_{BC} = (\textbf{F}_{m} \otimes \textbf{F}_{n})^{*}(\textbf{D})(\textbf{F}_{m} \otimes \textbf{F}_{n})  
\end{equation}
\section{Block Circulant symmetry in case of $n_{bpm} =  2n_{corrector}$}
The concept of DFT-based decomposition of ORM can also be extended to the matrices where $n_{bpm} =  2n_{corrector}$ and the orbit response matrix can be arranged in the form of two Circulant blocks each of dimension $n$ as
\begin{equation}
    \label{eq34}
    \textbf{R}=
\begin{bmatrix}
        a_0&a_1&a_2&a_3&.&.&a_{n-1} \\
        a_{n-1}&a_0&a_1&a_2&.&.&a_{n-2}\\
        .&.&.&.&.&.&. \\ 
         a_1&a_2&a_3&a_4&.&.&a_0 \\
        b_0&b_1&b_2&b_3&.&.&b_{n-1} \\
        b_{n-1}&b_0&b_1&b_2&.&.&b_{n-2} \\
        .&.&.&.&.&.&. \\ 
        b_1&b_2&b_3&b_4&.&.&b_0 \\
\end{bmatrix}
\end{equation}
and can be diagonalised by the DFT of the first rows of each block as 
\begin{equation}
    \label{eq35}
    \textbf{R} = (\frac{1}{n})\textbf{F}^{*}_c\mathbf{\Lambda}_c\textbf{F}
\end{equation}
where \textbf{F} is the standard Fourier matrix of size $n$. $\textbf{F}_{c}$ is a  $2n \times 2n$ matrix constructed using the standard Fourier matrix of size $n$ as 
\begin{equation}
    \label{eq36}
    \textbf{F}_{c}=
\begin{bmatrix}
        \textbf{F}&\textbf{0} \\
        \textbf{0}&\textbf{F}
\end{bmatrix}
\end{equation}
and $\mathbf{\Lambda}_c$ is a rectangular matrix composed of two diagonal block matrices as
\begin{equation}
    \label{eq37}
    \mathbf{\Lambda}_c =
\begin{bmatrix}
        \sigma_{a,0}&0&0&0&.&.&0 \\
        0&\sigma_{a,1}&0&0&.&.&0\\
        .&.&.&.&.&.&. \\
         0&0&0&0&.&.&\sigma_{a,n-1} \\
        \sigma_{b,0}&0&0&0&.&.&0 \\
        0&\sigma_{b,1}&0&0&.&.&0\\
        .&.&.&.&.&.&. \\
         0&0&0&0&.&.&\sigma_{b,n-1} \\
\end{bmatrix}
\end{equation}
where $\sigma_{a,i}$ and $\sigma_{b,i}$ are the Fourier coefficients of the first rows of each block of \textbf{R}, respectively.
The final decomposition can be written as 
\begin{equation}
    \label{eq38}
    \textbf{R}=
\begin{bmatrix}
        \textbf{F}^*&\textbf{0} \\
        \textbf{0}&\textbf{F}^*
\end{bmatrix}
\begin{bmatrix}
        \mathbf{\Lambda}_a\\
        \mathbf{\Lambda}_b
\end{bmatrix}
\begin{bmatrix}
        \textbf{F}
\end{bmatrix}
\end{equation}
and the pseudo-inverse of \textbf{R} can be calculated as 
\begin{equation}
    \label{eq39}
    \textbf{R}^{+}=
    \begin{bmatrix}
        \textbf{F}^*
\end{bmatrix}
\begin{bmatrix}
        \mathbf{\Lambda}_a^{+}&\mathbf{\Lambda}_b^{+}
\end{bmatrix}
\begin{bmatrix}
        \textbf{F}&\textbf{0} \\
        \textbf{0}&\textbf{F}
\end{bmatrix}
\end{equation}
Moreover, the SVD singular values of the matrix \textbf{R} also have a direct relation to the Fourier coefficients of individual blocks as
\begin{equation}
    \label{eq40}
    \centering
    s_{i} = \sqrt{|\sigma_{a,i}|^{2} + |\sigma_{b,i}|^{2}}
\end{equation}


\end{document}